\documentclass{ws-ijmpa}

\begin{document}

\markboth{P.~Moskal -- COSY-11 Collaboration}
{$\eta$ and $\eta^{\prime}$ mesons production at COSY-11}

%
\catchline{}{}{}{}{}
%

\title{ $\eta$ AND $\eta^{\prime}$ MESONS PRODUCTION AT COSY-11}

\author{\footnotesize P.~Moskal$^{\star,\%}$$^,$\footnote{E-mail address: 
p.moskal@fz-juelich.de}~, 
H.-H.~Adam$^{\#}$, 
A.~Budzanowski$^{\$}$,
E.~Czerwi\'nski$^{\star}$,
R.~Czy{\.z}ykiewicz$^{\star,\%}$,
D.~Gil$^{\star}$,
D.~Grzonka$^{\%}$, 
M.~Janusz$^{\star,\%}$, 
L.~Jarczyk$^{\star}$, 
B.~Kamys$^{\star}$, 
A.~Khoukaz$^{\#}$, 
P.~Klaja$^{\star,\%}$, 
J.~Majewski$^{\star,\%}$, 
W.~Oelert$^{\%}$,  
C.~Piskor-Ignatowicz$^{\star}$, 
J.~Przerwa$^{\star,\%}$, 
J.~Ritman$^{\%}$, 
B.~Rejdych$^{\star}$,
T.~Ro\.zek$^{+}$, 
T.~Sefzick$^{\%}$, 
M.~Siemaszko$^{+}$, 
J.~Smyrski$^{\star}$, 
A.~T\"aschner$^{\#}$, 
P.~Winter$^{\times}$,  
M.~Wolke$^{\%}$, 
P.~W\"ustner$^{\%}$, 
W.~Zipper$^{+}$ 
}

\address{
$^{\star}$Institute of Physics, Jagellonian University, Cracow, Poland\\ 
$^{\%}$IKP and ZEL, Forschungszentrum J\"ulich, J\"ulich, Germany\\ 
$^{\#}$IKP, Westf\"alische Wilhelms-Universit\"at, M\"unster, Germany\\
$^{\$}$Institute of Nuclear Physics, Cracow, Poland\\ 
$^{+}$Institute of Physics, University of Silesia, Katowice, Poland\\
$^{\times}$Department of Physics, University of Illinois at Urbana-Champaign, Urbana, IL 61801 USA\\
}

\maketitle


\begin{abstract}
The low emittance and small momentum spread of the proton
and deuteron beams  of the Cooler Synchrotron COSY
combined with the high mass resolution of the COSY-11
detection system permit to study the creation of mesons
in the nucleon-nucleon interaction
down to the fraction of MeV
with respect to the kinematical threshold.
At such small excess energies, the ejectiles possess low relative
momenta and are predominantly produced with the relative angular momentum equal to zero.
Taking advantage of these conditions we have performed
investigations aiming to determine the mechanism of the production 
of  $\eta$ and $\eta^{\prime}$ mesons in the collision of hadrons as well as
the hadronic interaction of these mesons 
with nucleons and nuclei.
In this proceedings we address the ongoing studies of 
the spin and isospin dependence for the production
of the $\eta$ and $\eta^{\prime}$ mesons
in free and quasi-free nucleon-nucleon collisions.

New results on the spin observables for the $\vec{p}p\to pp\eta$ reaction,
combined with the previously determined total cross section isospin dependence,
reveal a statistically significant indication that the excitation 
of the nucleon to the $S_{11}(1535)$
resonance, the process which intermediates the production of the $\eta$ meson 
in the nucleon-nucleon interactions, is predominantly due to the
exchange of the $\pi$ meson between the colliding nucleons. 

\keywords{meson-nucleon interaction; near threshold meson production.}
\end{abstract}

\section{Introduction}  
In the low energy limit, for energies lower than the  $\Lambda_{QCD}$ parameter\cite{newPDG},
in the domain where the strong coupling constant is large, 
there exists no clear description of the strong interaction since both 
quark-gluon and hadron degrees of freedom become relevant.
Therefore, in order to understand the phenomena governed by the strong forces 
in this non-perturbative regime of QCD
still a lot of experimental and theoretical effort is required.
In this energy regime, investigations of the production 
and decay of hadrons (objects owing their existence to the strong forces),
deliver information needed to deepen our knowledge about 
the strongly coupled QCD, where the perturbative approach is not possible.

Here we will focus on the studies of the production of
the $\eta$ and $\eta^{\prime}$ mesons emphasising such aspects like 
the production mechanism of these mesons and their interaction
with nucleons.  We will stress mainly results obtained at the COSY-11 
facility\cite{brauksiepe} operating at the cooler synchrotron COSY\cite{meier}.
 Yet,
whenever it will be possible, 
investigations of the COSY-11 group will
be presented in the broader context 
together with the relevant 
data obtained at other facilities. 
We hope to be able to demonstrate that, although 
there are always many possible interpretations of the determined observables,
the combination of
the energy dependence of the total cross section with 
its differential 
distribution and its 
spin and isospin dependencies, gathered during the decade of measurements,
permit now conclusive statements about the studied phenomena. 

\section{Advantages of the threshold kinematics}
The COSY-11 facility is designed for  studies of the mesons and hyperons
production in the nucleon-nucleon, nucleon-deuteron, 
and deuteron-deuteron collisions
near the kinematical threshold. 
For the details concerning the detection 
system\cite{brauksiepe,nimjurek}
as well as the methods of particle identification\cite{pawel}, absolute 
normalisation\cite{nim2001} or multidimensional acceptance corrections\cite{prc}
the interested reader is referred
to the quoted publications,
where the facility was described in a comprehensive way. 

Exactly at the reaction threshold 
all ejectiles are at rest in the center of mass system.
Therefore, in the case of the fixed target experiments,
due to the momentum conservation,
outgoing particles are confined in the laboratory
in a small cone centered around the beam line
and can be detected  by means of relatively small
detectors. In practice, it means that 
a full space phase 
coverage can be achieved even 
when using  magnetic spectrometers 
which are usually limited by a small geometrical acceptance.
This feature allows to combine a precise momentum reconstruction 
of the outgoing particles
with an effectively large detection efficiency.\\
In the case of the studies of  short-lived 
mesons, measured indirectly 
via the missing mass technique a very important 
advantage is that the  
missing mass resolution  due to the uncertainties 
of the reconstruction of the ejectiles' momenta tends to zero at threshold.
In addition, the smearing
of the missing mass distribution caused by the beam momentum spread 
is also narrowing with
decreasing beam momentum and it reaches 
its minimum at threshold\cite{erykproposalwidth}.
We demonstrated empirically that the missing mass 
resolution 
is approximately proportional to the 
square root of the excess energy\cite{jureketa}. 
Hence, we can benefit thoroughly from the threshold kinematics as far as the acceptance,
resolution of the missing mass reconstruction as well as a signal-to-background ratio are concerned.
Recently at COSY by means of the GEM\cite{machnereta05}
 setup the mass of the $\eta$ meson was determined with precision\cite{gemmass}, 
and in the near future, taking advantage of the threshold kinematics, 
using the developed monitoring methods\cite{nim2001} and the stochastically 
cooled proton beam of COSY
we will use the COSY-11 setup to measure directly
the natural width of the $\eta^{\prime}$ meson with an
experimental resolution of about 0.2~MeV\cite{erykproposalwidth}. 
\begin{figure}[h]
\vspace*{-4 mm}
\parbox{0.49\textwidth}{
\includegraphics[width=0.4\textwidth,angle=-90]{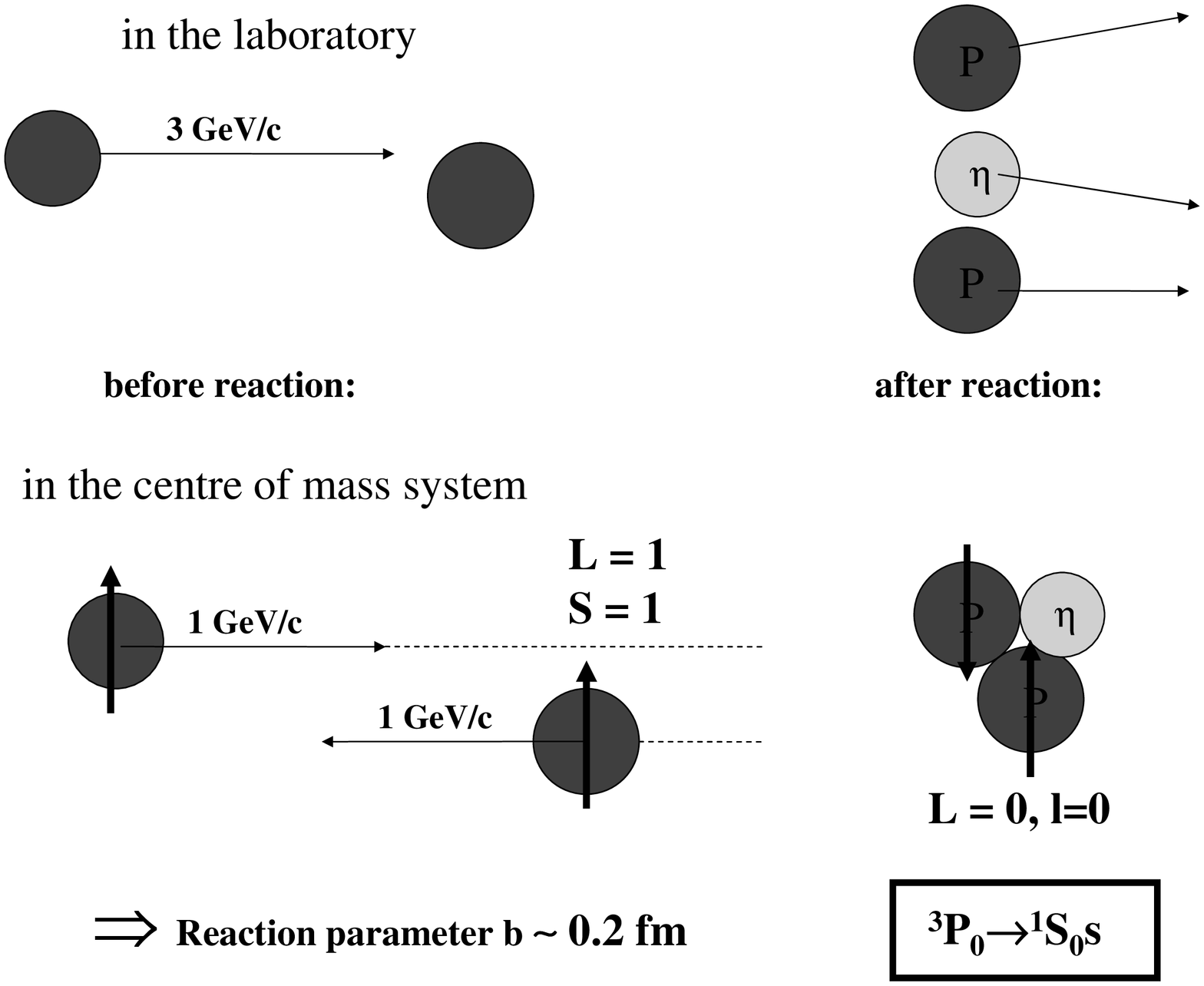}
}
\parbox{0.49\textwidth}{
\includegraphics[width=0.5\textwidth]{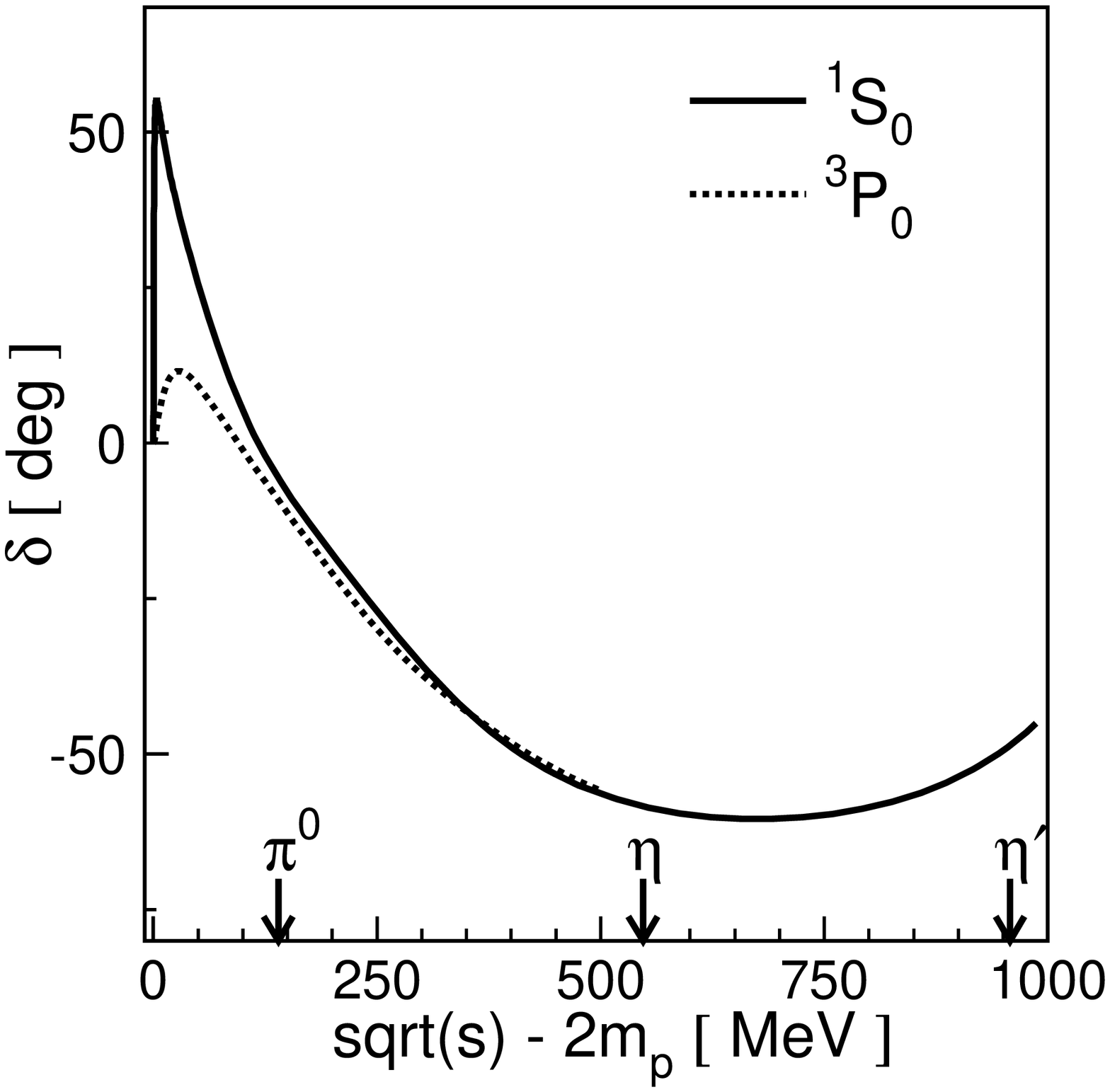}
}
{\caption{ 
({\bf Left}):  Schematic view of the $pp\to pp\eta^{\prime}$ process at threshold.
({\bf Right}): The $^1\mbox{S}_0$ and $^3\mbox{P}_0$
 phase-shifts of the nucleon-nucleon potential shown versus the
 centre-of-mass kinetic energy available in the proton-proton system. The
 values have been extracted from the SAID data base~\protect\cite{arndt3005}.
 \label{fig1}
}}
\vspace*{-4 mm}
\end{figure}
On the theoretical side
the most crucial facilitation and attractiveness when interpreting  meson production
at the vicinity of the threshold is the fact that the relative angular momenta larger than
$l = 0$ play no role due to the short range of the strong
interaction and small relative momenta of the produced particles.
Due to
the conservation laws and the Pauli excluding principle
for many reactions there is only one possible angular momentum and spin
orientation for the incoming and 
outgoing particles\footnote{The Pauli
  principle for the nucleon-nucleon system implies that
   $  (-1)^{{\scriptsize L}+{\scriptsize S}+{\scriptsize T}} = -1,$
  where $L$, $S$, and $T$ denote angular momentum, spin, and isospin of the nucleon
  pair, respectively. 
For the $NN\to NN X$ reaction
the conservation of the basic quantum numbers 
requires\cite{christoph} that
$  (-1)^{(\Delta S + \Delta T)}~=~\pi_{X} \ \ (-1)^{l}, $
where $\pi_{X}$ describes the intrinsic parity of meson X,
$\Delta S$ denotes the change in the spin,
and $\Delta T$ in isospin, between the initial and final $NN$ systems.
For more circumstantial discussion 
on the partial waves contribution 
in the production of 
various mesons in the collisions of nucleons an interested
reader is referred to references~\refcite{hab,deloff,christoph}.}.
Thus, the production of neutral mesons with negative parity -- as
pseudoscalar or vector mesons -- may proceed in the proton-proton collision
near threshold only via the transition between $^3\mbox{P}_{0}$ and
$^1\mbox{S}_0\mbox{s}$ partial waves.
This means that only the collision of protons with relative angular momentum equal
to $1\,\hbar$
may lead to the production of such mesons.
A situation which is pictorially demonstrated in
figure~\ref{fig1}(left).\\
Thus, basing on  
general conservation rules one can deduce that in the case of  
$\eta$ and $\eta^{\prime}$ production in  proton-proton collisions
the dominant transition is the one between $^3\mbox{P}_{0}$ and
$^1\mbox{S}_0\mbox{s}$ partial waves. 
Before coming to the experimental results let us still examine 
phase-shifts of the nucleon-nucleon potential (see figure~\ref{fig1}(right))
for the partial waves involved in the production process.
The $^3\mbox{P}_0$ phase-shift
variation 
in the vicinity of
the threshold for mesons heavier than $\pi^0$ is very weak,
and hence we expect that the interaction of nucleons before 
the act of a primary production  will not introduce a significant
energy dependence to the cross section excitation 
function\footnote{
Authors of reference~\refcite{hanhart176} have demonstrated that the
reduction factor due to the influence of the $NN$ initial state
interaction (ISI) can be  
estimated from the
phase-shifts and inelasticities only.
At the threshold for $\pi$ meson production ISI makes almost no distortion
since the reduction factor  is close to unity. This is because  at
this energy the inelasticity is still nearly 1 and the $^3\mbox{P}_0$
phase-shift is close to zero (see figure~\ref{fig1}(right).
However, at the $\eta$ threshold, where the phase-shift approaches its
minimum, the proton-proton ISI diminishes the total cross section already by
a factor of five.}.

Due to the large momentum transfer between colliding protons needed to create a
meson, 
the primary production amplitude 
is also only weakly energy
dependent in the excess energy
range of a few tens of MeV.
Directly at threshold, where all ejectiles are at rest in the centre-of-mass
frame, the momentum transfer is equal to the centre-of-mass momentum of the
interacting nucleons. In the case of the $\eta^{\prime}$ meson production 
it is equal to about 1~GeV/c~$\approx$~5~fm$^{-1}$, 
which according to the Heisenberg uncertainty relation
implies a 
distance of about 0.2~fm probed by the $NN \rightarrow NN \eta^{\prime}$ 
reaction at threshold. 
In contrast, the typical range of the strong nucleon-nucleon
interaction at low energies determined by the pion exchange may exceed a
distance of a few femtometers and hence 
is by one order of magnitude larger than the spatial size of 
the range where the production occurs.
Thus, in analogy to the Watson-Migdal approximation for two-body
processes~\cite{watson1163} the complete transition matrix element of
the production process may be factorized into
the total short range production amplitude ($M_{0}$),
and the interaction among particles in the exit ($M_{FSI}$) and initial ($F_{ISI}$) channels.
In contrary to the weak energy dependence of  
$M_{0}$ and $F_{ISI}$ we expect a strong variation of $M_{FSI}$
when the excess energy changes by tens of MeV.
This is due to the rapid changes of the phase-shifts for the $^1\mbox{S}_0\mbox{s}$ 
partial wave as it is demonstrated in figure~\ref{fig1}(right).
Therefore, near threshold, 
the shape of excitation functions for the 
total cross section 
for meson production in the collision of nucleons
will be predominantly due to the final state interaction between 
outgoing nucleons convoluted with the variation with energy of the phase space volume
available for the reaction.

\section{Signals from final state interaction}
\begin{figure}[h]
\vspace*{-5 mm}
\parbox[c]{0.40\textwidth}{
\centering
\includegraphics[width=0.4\textwidth]{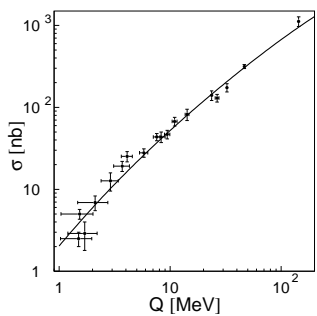}
}\hfill
\parbox{0.5\textwidth}{\caption{ 
Total cross section for the $pp\to pp\eta^{\prime}$ reaction as a function
of the excess energy. The data were obtained using 
COSY-11\protect\cite{etapalfons,etapc11pl,etapc11prl},
SPESIII~\protect\cite{hibou41} and DISTO~\protect\cite{etapdisto} experimental facilities.
The line shows the result of the parametrisation described in the text.
The parameters $\epsilon$ and the normalisation constant 
have been fixed by minimising the $\chi^2$ value.
The fit lead to the values of $\epsilon~=~0.62~\pm~0.13$~MeV and $const~=~84~\pm~14~mb$.
\label{fig2}
}}
\end{figure}

\begin{figure}[h]
    \parbox{0.33\textwidth}{
      \includegraphics[width=0.34\textwidth]{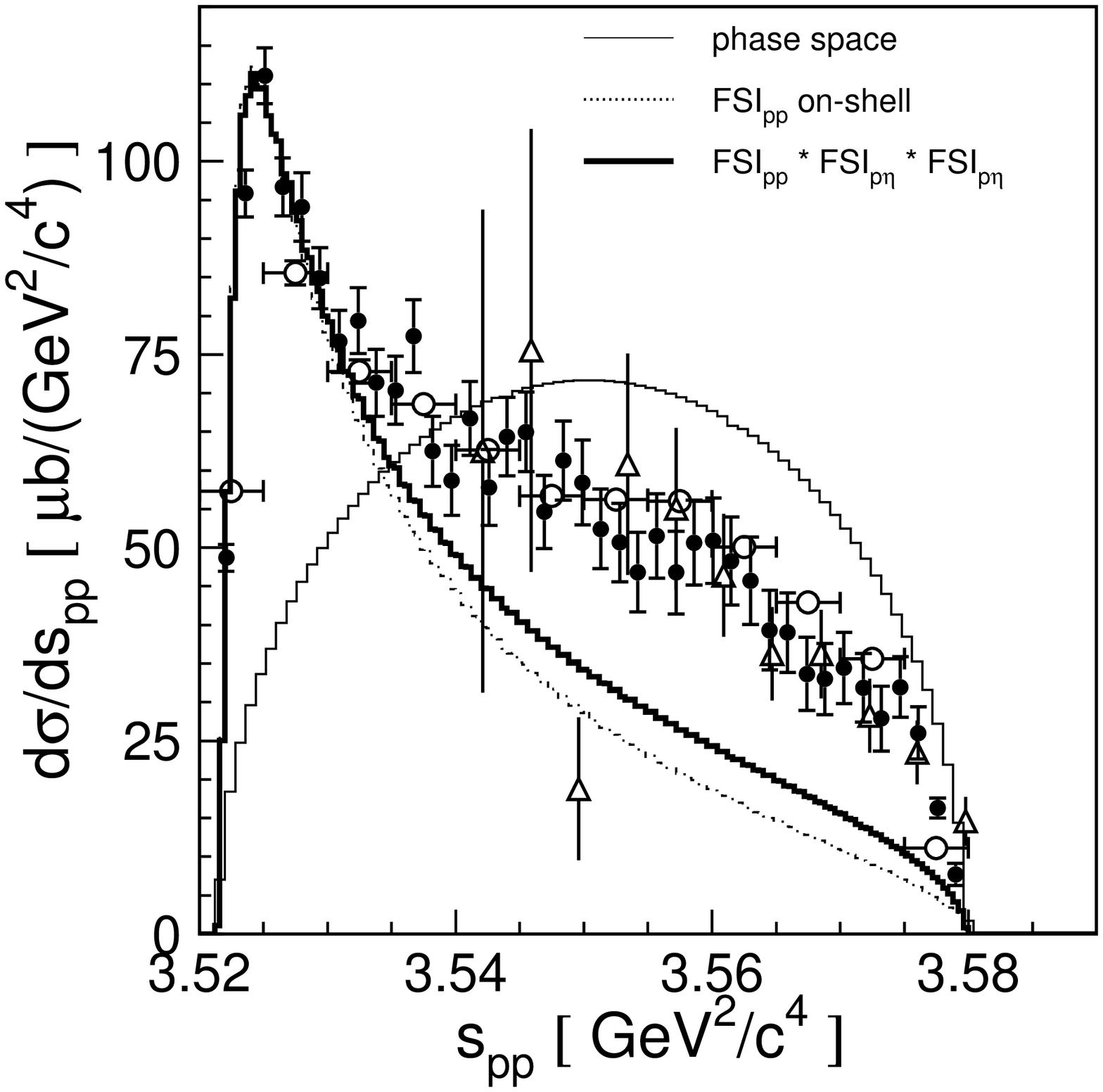}
    }\hfill
    \parbox{0.33\textwidth}{
      \includegraphics[width=0.34\textwidth]{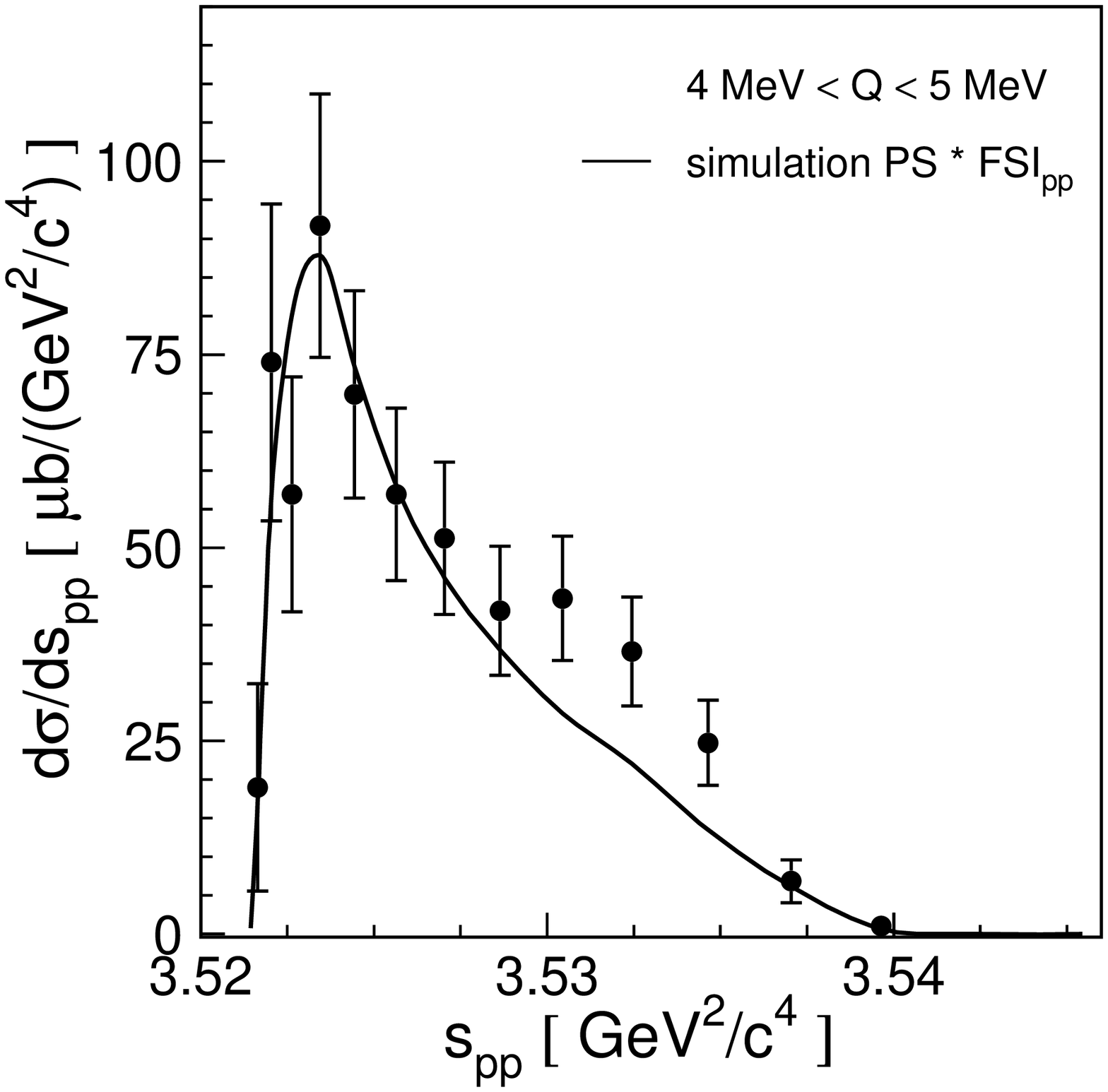}
    }\hfill 
    \parbox{0.33\textwidth}{
      \includegraphics[width=0.31\textwidth]{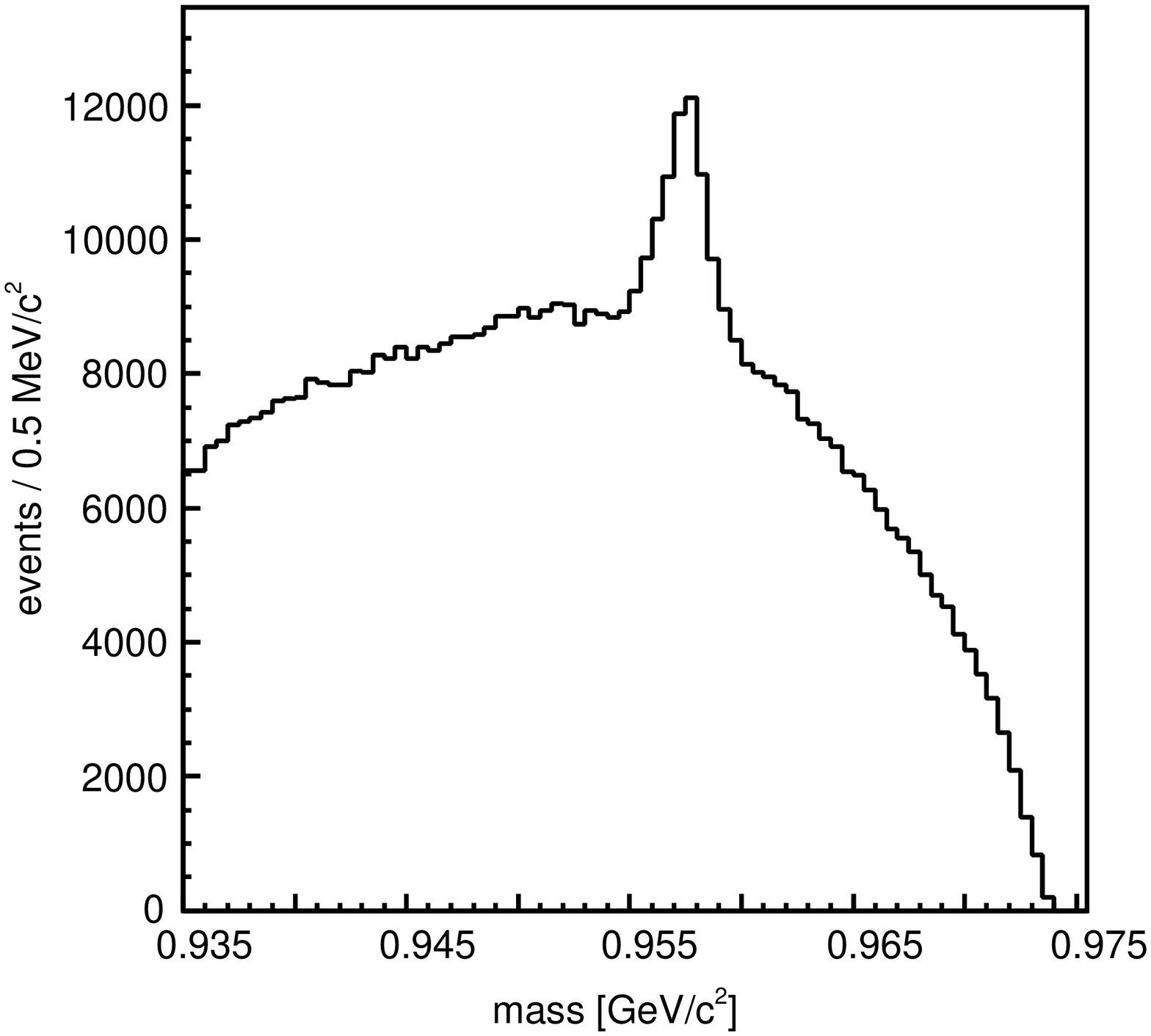}
    } 
    \caption{{\bf Left:}
         Distributions of the square of the proton-proton~($s_{pp}$)
          invariant mass
          determined experimentally
          for the $pp  \to pp\eta$ reaction at the excess energy of Q~=~15.5~MeV
          by the COSY-11 collaboration (closed circles),
          at Q~=~15~MeV by the TOF collaboration (open circles)\protect\cite{TOFeta},
          and at Q~=~16~MeV by PROMICE/WASA (open triangles)\protect\cite{calen190}.
    The integrals of the phase space weighted by
    the square of the proton-proton on-shell
    scattering amplitude~FSI$_{pp}$(dotted line), and by the product of FSI$_{pp}$ and
    the square of the proton-$\eta$ scattering amplitude~(thick solid line)
    have been normalized arbitrarily at small values of $s_{pp}$.
    The expectation for homogeneously populated phase space
    is shown as thin solid curve\protect\cite{prc,hab}.
    {\bf Center:}
      Distribution of the square of the proton-proton invariant mass from the $pp\to pp\eta$ reaction
      measured at COSY-11 for the excess energy range
      4~MeV~$\le$~Q~$\le$~5~MeV\protect\cite{prc}.
      The superimposed line shows the result of simulations assuming that
      the phase space population is determined exclusively by the on-shell interaction
      between outgoing protons\protect\cite{prc,hab}.
    {\bf Right:}
      Missing mass distribution for the $pp\to ppX$ reaction measured using 
      the COSY-11 facility at an excess energy
      of Q~=~15.5~MeV above the threshold for the $\eta^{\prime}$ meson production.
    \label{fig3}
    }
  \end{figure}

Now we can confront results of considerations
carried out in the previous section
with the experimental data. 
Figure~\ref{fig2} presents a total cross section for the 
$pp\to pp\eta^{\prime}$ reaction as a function of the excess energy.
The solid line superimposed on the data indicates calculations
of the total cross section performed employing factorisation
of the total production matrix element $|M_{pp \rightarrow pp X}|$ into
the short range primary amplitude $|M_0|$ and the  initial and final state interaction
$|M_{pp \rightarrow pp X}|^2 \approx |M_{FSI}|^2 \cdot |M_0|^2 \cdot F_{ISI}$.
The proton-proton FSI effects have been taken into account
according to the
model developed by F{\"a}ldt and Wilkin\cite{faldt209,faldt2067},
which allows to express the total cross section energy dependence for a $NN
\rightarrow NN\,Meson$ reaction by a closed analytical formula\cite{hab}:
$\sigma \;=\;
  const \cdot \frac{V_{ps}}{\mbox{F}} \cdot
    {\left(1\;+\;\sqrt{1\,+\,\frac{\mbox{\scriptsize Q}}{{\displaystyle \epsilon}}}\right)^{-2},}$
where Q stands for the excess energy, F denotes the flux factor\cite{bycklingkajantie}
and $V_{ps}$ corresponds to the volume of the available phase space\cite{hab}. 
The parameter $\epsilon$ and the normalisation constant 
have to be settled from the data.
Figure~\ref{fig2} demonstrates that the data can indeed be very well described
under the above discussed Ansatz. We can conclude also, that the interaction
of the $\eta^{\prime}$ meson with the proton, which was neglected in the calculations,
is too weak to manifest itself visibly within the statistical error bars\cite{swave}.
Interestingly, when using the same model and calculating the shape for the 
excitation function of the $\eta$ meson production we underestimated the
data at the very threshold by about a factor of 
two\cite{hab,prc}~\footnote{Recently an even larger enhancement has been observed
in the case of the $K^+K^-$ pair production\cite{winter23,KKwalter}. 
It cannot be excluded that the  effect is due to the 
strong $K^+K^-$ interaction. The interpretation is however still open. }.
The difference could be 
explained when extending the factorisation by the proton-$\eta$ interaction,
however doing so we  fail to describe the invariant mass distributions
where the discrepancy is even more pronounced as can be clearly observed
in figure~\ref{fig3}(left). 
In order to explain the structure observed in the invariant mass spectra 
Nakayama and collaborators\cite{nakayama2} suggest a
contribution from higher partial waves to the production process.
In fact an admixture of the $^1S_{0}\to ^3\!\!P_{0}s$ transition to the main
  $^3P_{0}\to ^1\!\!S_{0}s$ one results in the
  very good agreement with the experimental points 
in the invariant mass spectra\cite{nakayama2}.
However, at the same time, this conjecture leads
  to strong discrepancies in the shape of the excitation function\cite{nakayama2,hab}.
Till now there exists no consistent picture allowing for the simultaneous 
explanation  of the excitation function and invariant mass distributions 
for the $\eta$ meson production. 
The enhancement is visible also at Q~=~4.5~MeV~(see figure\ref{fig3}(center))\cite{prc}, where the
contribution of the higher partial waves is quite improbable\cite{review}.
We deem this as an indication in favour of the hypothesis 
that the effect 
is caused by the proton-$\eta$ interaction rather than by  higher partial waves.
 In order to shed new light on these investigations
we have conducted a high statistics measurement of the $\eta^{\prime}$ meson 
creation at an excess energy of Q~=~15.5~MeV.
 Our purpose is to determine an invariant 
mass distribution of the $p\eta^{\prime}$ system at exactly the same value of excess
energy as it was done in the case of the $\eta$ meson. If for the $\eta^{\prime}$ meson 
case a similar enhancement appears it will indicate that its origin 
cannot be assigned to the meson-proton interaction and hence it would strengthen
the hypothesis suggesting a significant contribution from higher partial waves\cite{nakayama2}.
On the other hand, if the enhancement disappears
this will raise the 
 confidence to the hypothesis that the observed bump is due to the 
proton-$\eta$ interaction acting in the $pp\eta$ system. 
The data are  being analyzed,
and presently as a herald of the forthcoming invariant mass distribution
we show
a missing mass spectrum~(figure~\ref{fig3}(right)) where a clear signal with 
about 17000 events corresponding to the $pp\to pp\eta^{\prime}$ reaction is clearly 
visible.

\section{The power of analysing power - $\eta$ production with polarized beam}
  A precise data set\cite{prc,hibou41,bergdolt2969,chiavassa270,calen39,calen79}
  on the total 
  cross section of the $\eta$ meson production
  in the $pp\to pp\eta$ reaction allowed to conclude that the reaction proceeds
  through the excitation of one of the protons to the $S_{11}(1535)$ state which 
  subsequently deexcites via emission of the $\eta$ meson.  The crucial observations
  were a large value of the absolute cross section (forty times larger than for the 
  $\eta^\prime$ meson) and  isotropic distributions\cite{prc,TOFeta} of the angle of the
  $\eta$ meson emission in the reaction center-of-mass system. In practice, in the meson
  exchange picture the excitation of the intermediate resonance can be induced
  by the exchange between the nucleons of any of the pseudoscalar or vector ground state mesons.
  Based only on the total cross section and its dependence on the excess energy 
  it was however impossible to falsify or confirm any of the proposed 
  hypothesis. 
  In fact due to the negligible variation of the production 
  amplitude in the range of few tens of MeV the full information 
  available from the excitation function is reduced to a single number\cite{bernard259}.

  Theoretical collaborations\cite{germond308,laget254,faldt427,moalem445,vetter153,nakayama,alvaredo125,batinic321}
  reproduce the magnitude of the total cross
  section, though their models differ significantly 
  as far as the relative contributions
  from the exchange of various  mesons  are concerned.
  The ambiguity was solved partially by the determination of 
  the isospin dependence of the total cross section by the WASA/PROMICE
  collaboration\cite{calen2667}. From the comparison of the $\eta$ meson 
  production in proton-proton and proton-neutron reactions
  it could be inferred that $\eta$ is by a factor of 12 more copiously produced 
  when the total isospin of the nucleons is equal to zero than when it is equal to one.
  As a  consequence only  an isovector mesons exchange is conceivable 
  as responsible for such a strong isospin dependence. It was a large step forward
  but still the relative contributions of $\rho$ and $\pi$ mesons
  remained to be disentangled. The spin averaged observables are 
  consistent with the calculations based upon $\rho$ meson exchange 
  dominance\cite{faldt427} as well as upon $\pi$ meson 
  exchange dominance\cite{nakayama2,nakayama}.  
  \begin{figure}[h]
   \vspace*{-3 mm}
   \parbox{0.26\textwidth}{
     \includegraphics[width=0.29\textwidth]{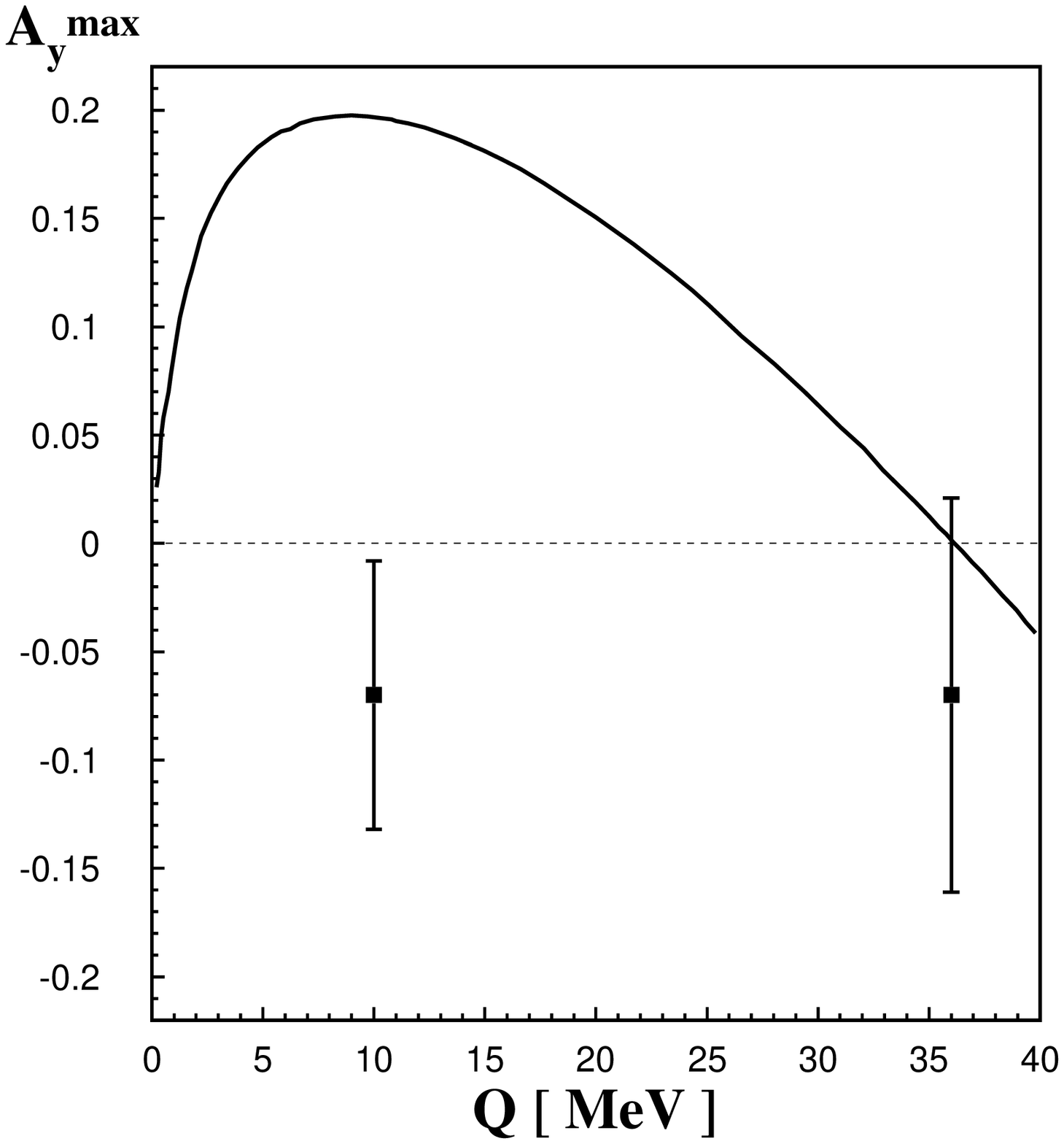}
   }
   \parbox{0.27\textwidth}{
     \includegraphics[width=0.29\textwidth]{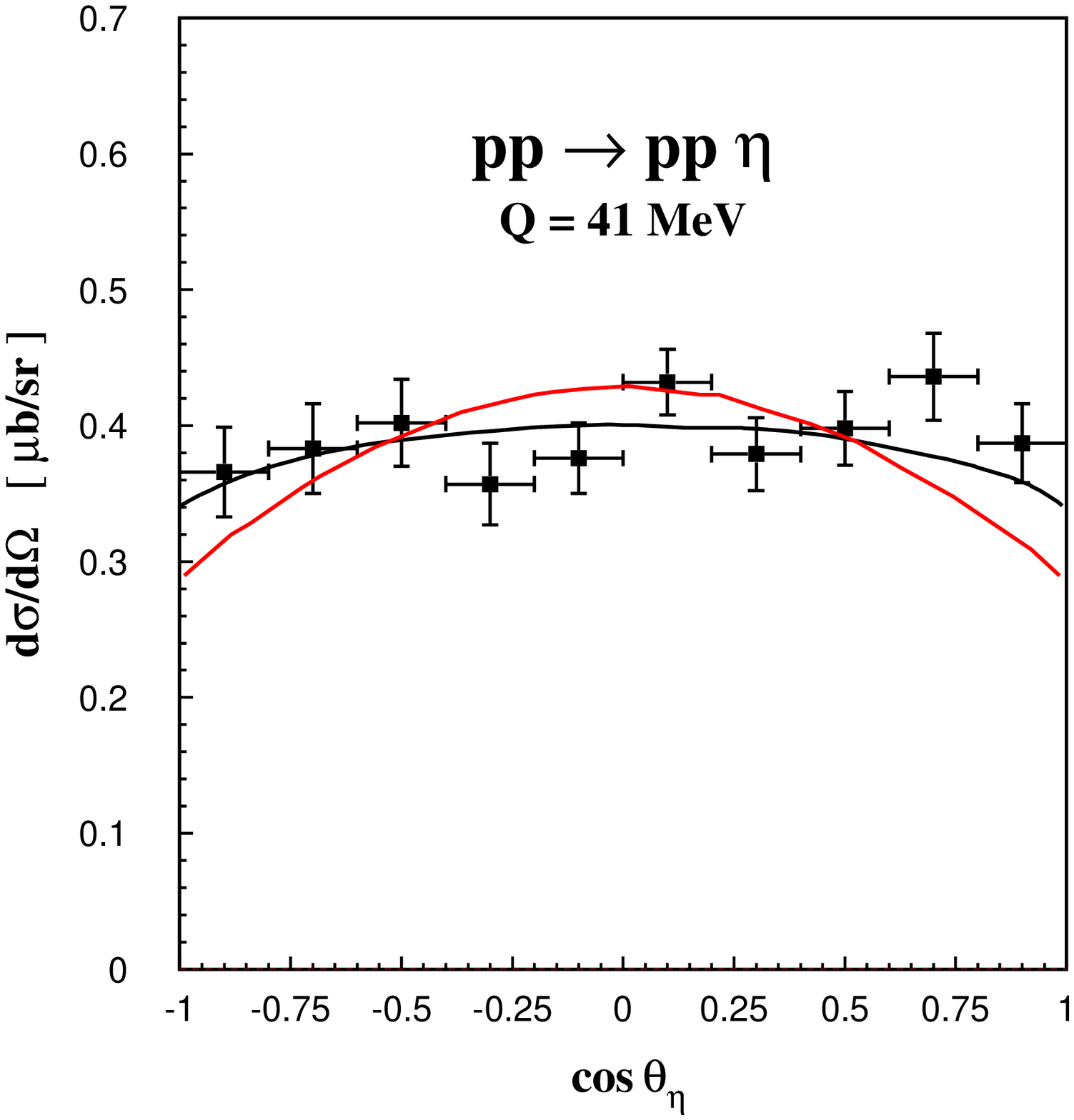} 
   }\hfill
   \parbox{0.44\textwidth}{
     \caption{{\bf Left}: Amplitude of the analysing power 
       as predicted upon vector dominance model\protect\cite{faldt427} (line) 
       and as measured by the COSY-11\protect\cite{czyzykmeson06} (data).
     {\bf Right}:
       Center-of-mass angular distribution of the $\eta$ meson emission.
       Data were measured by the COSY-TOF\protect\cite{TOFeta}.
       Less and more bend lines correspond 
       to $\pi$ and $\rho$ dominance, respectively\protect\cite{nakayama}.
     \label{fig4}
   }
 }
 \vspace*{-3 mm}
 \end{figure}
  Yet, the conclusions drawn for the angular dependence of the beam analysing
  power are significantly different depending on whether $\pi$ or $\rho$ meson
  dominance is assumed as a leading mechanism for exciting  one of the colliding 
  nucleons\cite{nakayama2,nakayama}. 
  Encouraged by the discovery potential given by the contradicting predictions
  we have performed an experiment aiming to determine the angular dependence of the 
  analysing power for the $pp\to pp\eta$ reaction. After a successful test run\cite{winter251}
  we have conducted measurements at excess energies of Q~=~10~MeV and Q~=~36~MeV\cite{czyzyk1}.
  As a result we have established that the analysing powers 
  for both excess energies are consistent with zero.  
  The $\chi^2$ analysis  excludes correctness of the assumption 
  about a pure vector meson dominance ($\rho$ exchange)
  with the significance level larger than four standard deviations,
  and provides strong evidence for the correctness
  of the supposition 
  that the production of  $\eta$ mesons in  nucleon
  nucleon collision is dominated by the pion exchange. 
  Selected results are shown in figure~\ref{fig4}, and for 
  more details the reader is referred to a report of Czy{\.z}ykiewicz\cite{czyzykmeson06} 
  in these proceedings.

\section{Isospin dependence of the hadronic production of the $\eta^{\prime}$ meson}
In the preceding section based upon spin and isospin observables for the
$NN\to NN\eta$ reaction we deduced that 
in collisions of nucleons the $\eta$ meson 
is primordially created through the exchange of pion leading to the excitation 
of one of the nucleons to the $S_{11}(1535)$ state which subsequently decays
into the $\eta$ meson and a nucleon.
In the case of the $\eta^{\prime}$ meson our understanding of the process
is still much poorer and unsatisfactory. 
We attempt to apprehend this process   
since there are many indications that the  
wave function of the $\eta^{\prime}$ meson
comprises a significant gluonic component\cite{bass245,bass187},
distinguishing it from other ground state  mesons, 
and we hope that the comprehension of the
mechanism leading to the creation of the $\eta^{\prime}$ meson 
in collisions of hadrons may help
to determine its quark-gluon structure. 
A potentially
 large glue content of the $\eta^{\prime}$ and the dominant
flavour-singlet combination of its quark wave function may cause that the
dynamics of its production process in nucleon-nucleon collisions is
significantly different from that responsible for the production of other
mesons.
In particular, the $\eta^{\prime}$ meson can be efficiently created via a
``contact interaction'' from the glue which is excited in the interaction
region of the colliding nucleons\cite{bass286,bass118,bass348}.

At present the models can be confronted with the 
values of the total cross section only, and 
until now it has not been possible to satisfactorily estimate the relative
contributions of the nucleonic, mesonic, and resonance current to the
production process\cite{nakayama}.
Therefore, in order to disentangle 
the ambiguities it is mandatory to determine experimentally spin and isospin observables.

\begin{figure}[h]
  \vspace*{-3 mm}
  \parbox[c]{0.33\textwidth}{
     \includegraphics[width=0.33\textwidth]{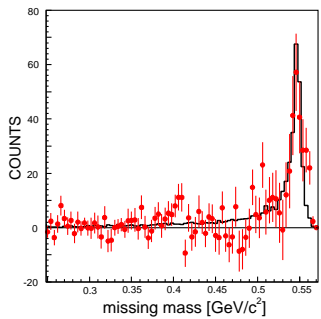}
  }\hfill
  \parbox[c]{0.59\textwidth}{
  \caption{ 
          Missing mass distribution of the quasi-free $pn\to pn X$
             process determined for $Q>0$ with respect to the
             $pn\to pn\eta$ reaction\protect\cite{jphys}.
             The sum for all $\Delta{Q}$ intervals is shown.
           The points denote the experimental data for $Q> 0$
             after subtraction of the multi-pion background.
             The superimposed solid line, normalised in amplitude to the data points,
             results from a Monte-Carlo simulation.
    \label{fig5}
 }}
  \vspace*{-3 mm}
\end{figure}

As a first step we have conducted measurements of the $pn\to pn\eta^{\prime}$ reaction
in order to establish an isospin dependence of the total cross section\cite{hadron,joannaleap}.
We expect that the result will help to judge about the isospin nature of the objects exchanged
which intermediate the production process.
On the other hand a very important theoretical result
is that regardless of whether it is a mesonic, nucleonic, or resonance  current
the contribution from the exchange of isovector mesons ($\rho$ or $\pi$)
is much larger compared to that of isoscalar
ones~($\omega$ or $\eta$)\cite{nakayama024001,wilkinjohansson,vadim024002}.
Hence, our conviction is that on the hadronic level the process should have a rather strong
isospin dependence, unless there is a fortuitous cancellation of the dominating 
amplitudes~\footnote{
This can only be verified from  spin observables in further studies which we 
intend to conduct at the WASA-at-COSY facility\cite{loiwinter,wasaatcosy}.}.
This  entails that if the
ratio  
$R_{\eta^{\prime}} = \frac{\sigma{(pn\to pn\eta^{\prime}})}{\sigma{(pp\to pp\eta^{\prime})}}$
 --~corrected for FSI and ISI distortions~--
will be found to be close to unity we will have an indication
that the $\eta^{\prime}$ is produced directly by  gluons.
On the way towards the determination of the value of $R_{\eta^{\prime}}$
by means of the COSY-11 facility
a test experiment of the $pn\to pn\eta$
reaction --~suspected  to have by at least a factor of
thirty larger cross section than the one
for the $pn\to pn\eta^{\prime}$ reaction~--
was performed\cite{hadron,jphys}.
  In this test measurement,
  using a beam of stochastically cooled protons
  and a deuteron cluster target,
  we have proven the ability of the COSY-11 facility to study
  the quasi-free creation of mesons via the $pn\to pn X$ reaction.
In figure~\ref{fig5} a clear signal originating from the quasi-free $pn\to pn\eta$ reaction 
is visible. In the data evaluation a spectator model was employed and the
background was subtracted according to the recently developed method\cite{jphys}.
The experimental distribution is fully consistent with expectations determined taking into 
account all effects introduced by the instrumentation system and 
the known  physical processes which particles undergo when passing through the 
detectors. In March this year we have completed data taking for the $pn\to pn\eta^{\prime}$
reaction. The obtained integrated luminosity is by factor of 50 larger than this
of the test measurement of the $pn\to pn\eta$ with the signal presented in figure~\ref{fig5}.
Presently the data are under analysis.
In order to measure the $pn \to pn Meson$ reactions 
we use a proton beam and a deuteron target.
The main conjecture of this approach is that the
bombarding proton interacts exclusively with one nucleon in the target nucleus
and that the other nucleon affects the reaction by providing a momentum
distribution to the struck constituent only~\footnote{A detailed description of the
application of this technique can be found e.g. in references~\refcite{c11quasi,stepaniak,review}.}.
In the case of the $\eta^{\prime}$ meson production,
due to the large centre-of-mass velocity ($\beta \approx 0.75$) with respect
to the colliding nucleons, a few MeV wide spectrum of the neutron kinetic
energy inside a deuteron is broadened by more than a factor of thirty.
Therefore,  to achieve an accuracy of the excess energy in
the order of few MeV it is important to reconstruct the four-momentum vector
of the interacting neutron on the event-by-event basis.
Such an accuracy is mandatory for  close-to-threshold studies
where the cross sections vary by few orders of magnitude within the range 
of excess energy of few tens of MeV\cite{review}.
For this purpose, the spectator proton is registered and its four momentum 
vector is reconstructed\cite{c11quasi,hab}.
Subsequently, energy and momentum conservation permit to determine
the four-momentum vector of the reacting neutron.\\
Finally, for the comparison of the results obtained from a quasi-free and  free reactions,
we need to make a second assumption namely 
that the matrix element for quasi-free meson production off a
bound neutron is identical to that for the free $pn \rightarrow p n\,Meson$
reaction.

\section{Is the spectator model valid?}
The title of this section constitutes a frequently expressed  concern
in the context of the investigations of the meson production
in the proton-neutron collisions  where the neutron is bound 
in the nucleus. A 
positive answer to 
it must be justified  
if we are to trust the results derived employing the spectator model.
Below we list a few simple arguments which build our confidence to this model
and more important we quote empirical results which confirm the validity 
of the spectator hypothesis on the few per cent level
and set limits of its applications.

(1) Based on  intuition from classical mechanics 
   the assumption {\em that only a hit nucleon takes part in the collision}
   is justified if the kinetic energy of a projectile is large
   compared to the binding energy of the  nucleus.
   Indeed, the deuteron is relatively weakly bound with a binding energy of
   $\mbox{E}_B \approx 2.2\,\mbox{MeV}$, which is more than three 
   orders of magnitude smaller 
   than the kinetic energy of
   the bombarding proton needed for the creation of the $\eta^{\prime}$ meson
   in the proton-neutron interaction.

(2)  In case of meson production off the deuteron, one can also
    justify the assumption of the quasi-free scattering with a geometrical argument,
    since the average distance between the proton and the neutron is
    in the order of $3\,\mbox{fm}$.
    Certainly, the other nucleon may scatter 
    the incoming proton and the outgoing meson.
    Yet, this nuclear processes referred to as a shadow effect and reabsorption, respectively,
    decrease the total cross section (e.g. for the $\eta$-meson production)
    by about a few per cent only\cite{chiavassa192,smith647}.

(3) Comparisons of the quasi-free and free angular distributions 
   for the $pp\to d\pi^+$ reaction done at the TRIUMF facility\cite{duncan4390}
   have confirmed the validity of both crucial hypotheses of the spectator model.
   It was demonstrated that the experimental spectator momentum distribution 
   conforms very well expectations based upon spectator model. 
   The experiment revealed also that the magnitude of the differential cross sections
   for the quasi-free $pp\to d\pi^+$ process agree on the few per cent level 
   with the free differential cross sections, thus proving also that the matrix
   element for the free and quasi-free process are equal 
   at least to this level level.
   It is important to note that  
   the energies of projectiles in this experiment were 
   few times lower than needed to produce $\eta$ or $\eta^{\prime}$ meson,
   and at higher energies the approximation should work even better.
   
(4) The WASA/PROMICE collaboration has compared quasi-free and free
   production cross sections for the $pp\to pp\eta$ reaction.
   As a result it was shown that within the statistical error bars
   there is no difference between the total cross section of the free 
   and quasi-free process. Thus confirming the validity of the assumption
   regarding the equality of the production matrix elements for free and bound nucleons.

(5) A dedicated empirical test of the first assumption 
  of the spectator model has been performed
  recently using the high acceptance COSY-TOF facility\cite{tofkuhlman}.
  The shape of the angular distributions for the quasi-free
  $np\to pp\pi^-$ and $pn\to pn$ reactions as well as the 
   form of the momentum distributions of the spectator  have been measured.
  Calculations based upon the hypothesis of the spectator model 
  yield results consistent with the experimental data with an
  accuracy better than 4\% up to 150~MeV/c of the Fermi momentum
  and with about 25\% up to a momentum of 300~MeV/c.
   
\section{Conclusion}
 Due to the limited space we could give only a brief account of a few chosen
 aspects of our investigations concerning the $\eta$ and $\eta^{\prime}$ physics.
 We did not mentioned e.g. the issues of the $\eta$ meson production in the few nucleon
 system\cite{jurekacta,hhadam} or
the search for a possible bound state of the $\eta$ meson with the nucleus 
of Helium\cite{jurekproposal}. 
 The interested reader is thus referred to the mentioned references.

\section{Acknowledgements}
We acknowledge the support of the
European Community-Research Infrastructure Activity
under the FP6 "Structuring the European Research Area" programme
(HadronPhysics, contract number RII3-CT-2004-506078),
of the FFE grants (41266606 and 41266654) from the Research Centre J{\"u}lich,
of the DAAD Exchange Programme (PPP-Polen),
of the Polish State Committe for Scientific Research
(grant No. PB1060/P03/2004/26), 
and of the EtaMesonNet.

\end{document}